\documentclass[]{aa}
\usepackage{amsmath,graphicx,natbib,aasjournals}

\bibpunct{(}{)}{;}{a}{}{,} 

\newcommand{\amax}{\mbox{a$_{\rm max}$}}
\newcommand{\amin}{\mbox{a$_{\rm min}$}}
\newcommand{\etal}{et~al.~}
\newcommand{\lsun}{\mbox{L$_{\sun}$~}}
\newcommand{\micron}{\mbox{$\mu$m}}

\newcommand{\msun}{\mbox{M$_{\sun}$~}}

\newcommand{\teff}{\mbox{$T_{\rm eff}$}}

\begin{document}

\title{The origin of crystalline silicates in the Herbig Be star HD100546 and 
       in comet Hale-Bopp}

\author{J.\,Bouwman\inst{1,2}
\and A.\,de Koter\inst{1}
\and C.\,Dominik\inst{1}
\and L.B.F.M.\,Waters\inst{1,3}}

\offprints{J. Bouwman}

\institute{Astronomical Institute "Anton Pannekoek", University of Amsterdam,
Kruislaan 403, 1098~SJ Amsterdam, The Netherlands
\and CEA, DSM, DAPNIA, Service d'Astrophysique, CE Saclay, 91191 Gif-sur-Yvette Cedex, France
\and Instituut voor Sterrenkunde, K.U. Leuven, Celestijnenlaan 200 B, B-3001
Heverlee, Belgi\"e}


\titlerunning{HD100546}
\authorrunning{Bouwman \etal}

\abstract{
We have investigated the spatial distribution, and the
properties and chemical composition of the dust orbiting HD\,100546. 
This system is remarkably different from other
isolated Herbig Ae/Be stars in both the strength of the mid-IR excess
and the composition of the circumstellar dust.
To explain spectral features and the amount of mid-IR dust emission
the presence of a component of
small ($<$ 10 $\mu$m) grains radiating at $\sim$200~K is required,
which is not seen in other well investigated Herbig Ae/Be systems.
This additional component is inconsistent with a uniform flaring disk model.
The fraction of intercepted stellar light that is absorbed
and re-emitted in the mid-IR is so large ($\sim$~70\,\%) that it
requires the disk to be more `puffed up' at about 10\,AU, where
the grains have $T \sim 200$ K. This may occur if a
proto-Jupiter clears out a gap at this distance allowing direct
stellar light to produce an extended rim at the far side of the
gap.
The other remarkable difference with other isolated Herbig Ae/Be systems
is the presence of a much larger mass fraction of the crystalline silicate forsterite
in the circumstellar dust.
We find that the mass fraction of crystalline silicates in
HD\,100546 increases with decreasing temperature, i.e. with
larger radial distances from the central star. This distribution
of crystalline dust is inconsistent with radial mixing models
where the crystalline silicates are formed by thermal annealing
above the glass temperature in the very inner parts of the
disk, and are subsequently transported outwards and mixed
with amorphous material.
We speculate that the formation and spatial distribution of the crystalline dust
may be linked to the formation of a proto-Jupiter in the disk around HD\,100546.
Such a proto-Jupiter could gravitationally stir the disk leading to a
collisional cascade of asteroidal sized objects producing small crystalline grains,
or it could cause shocks by tidal interaction with the disk which might produce
crystalline dust grains through flash heating.
As shown by \citet{hd100546malfait}, the infrared spectrum of HD\,100546 is
very similar to that of C/1995 O1 Hale-Bopp \citep{crovisier1997}.
Using an identical methodology, we have therefore also studied this
solar system comet. Both objects have an almost identical grain composition, but
with the important difference that the individual dust species in
Hale-Bopp are in thermal contact with each other, while this is not the case in HD\,100546.
This suggests that if similar processes leading to the dust composition as seen in HD\,100546
also occurred in our own solar system, that Hale-Bopp formed after the formation
of one or more proto-gas giants.
}

\maketitle

\section{Introduction}
Herbig Ae/Be stars (hereafter referred to as HAEBE stars) were first
described as a group by \citet{herbig1960}, in a study which was aimed at
finding intermediate mass young stars.  Many studies have since
confirmed the young pre-main-sequence (PMS) nature of HAEBE stars. 
Hipparcos parallaxes (e.g.  van den Ancker et al.  1998) allowed
accurate positions in the HR diagram to be obtained, and from comparison
to PMS evolutionary tracks masses of the order of 2 to 8 M$_{\odot}$ and
stellar ages measured relative to the birthline
of typically 10$^{6}$ to 10$^{7}$ yrs were found.  Direct
imaging at millimeter wavelengths of several Herbig Ae stars revealed
the presence of rotating flattened structures \citep{mannings1997,mannings2000}, 
believed to be the remnant of the accretion disk and the site of
on-going planet formation. It is not clear whether disks are also
common around more massive Herbig Be stars.

At infra-red (IR) wavelengths, the Infrared Space Observatory 
\citep[ISO][]{kessler1996} has obtained full 2-200 $\mu$m spectra of the
brightest HAEBE stars. 
These spectra show a wealth of detail concerning the thermal
emission of dust in the circumstellar environment. In a series of
papers, we have studied the dust emission from HAEBE stars
\citep{waelkens1996,hd100546malfait,malfait1999,mario_abaur,mario2001,abaur,
processing,HerbigOverview},
focusing on the mineralogical composition of the dust. These studies
indicate that a substantial modification of the dust composition from that in
the interstellar medium (ISM) occurs on time scales that are still poorly
constrained, but are generally less than 10$^{7}$ years. We mention
grain growth, the formation of crystalline silicates and of crystalline
H$_2$O ice, and of hydro silicates. These
modifications are important clues to the processes that eventually lead
to planet formation, and which can be compared to the records of planet
formation as found in solar system objects (meteorites, comets,
interplanetary dust particles). 

In this study we re-examine the ISO spectrum of HD\,100546, first
presented by \citet{waelkens1996}.
This
object has an exceptionally high fraction of crystalline silicates, and
its ISO spectrum shows a remarkable resemblance to that of the solar
system comet Hale-Bopp \citep{crovisier1997}. 
\citet{hd100546malfait} already suggested the
exciting possibility that we are witnessing the birth of an Oort cloud
with a multitude of cometary bodies being scattered by a (hypothesized)
giant planet. \citet{processing} analyzed the
10 $\mu$m silicate band of 14 HAEBE stars including HD\,100546, and found
that its mineralogical composition deviates substantially
from that of other HAEBE systems: in addition to a large fraction of
forsterite, the average grain size of the particles causing the 10
$\mu$m emission is larger.
Also, a lack of silica compared to forsterite was found. The latter can be
interpreted as a difference in the crystallization  process leading to the 
formation of forsterite.
We decided that a closer look at the
distribution of the different mineralogical components in terms of mass
and temperature is needed to better understand the nature of HD\,100546. 
We also analyze the ISO spectrum of Hale-Bopp, using the same analysis 
method as for HD\,100546. 
Previously, we have carried out a similar detailed analysis for AB\,Aurigae 
and HD\,163296 \citep{abaur} and we will compare these results
with those obtained here for HD\,100546. 
The observations of HD\,100546 used in this paper are taken from
\citet{hd100546malfait} and \citet{HerbigOverview}. 
The observations of Hale-Bopp were secured when the comet had a
heliocentric distance of 2.8~AU and are from \citet{crovisier1997}.

This paper is organized as follows: in Sect.~\ref{sect:hd-energy} we discuss 
the difference in dust composition and spectral energy distribution of HD\,100546 
compared to other HAEBE systems. In Sect.~\ref{sect:hd_modelling} we explain
the method to fit the spectrum. Sect.~\ref{sect:hd-results} describes our 
results for HD\,100546 and for Hale-Bopp. In Sect.~\ref{sect:hd-discussion} we 
compare our results to those obtained for AB\,Aur and HD\,163296, and discuss 
the implications for the formation process of the crystalline silicates
in HD\,100546. Sect.~\ref{sect:hd-summary} summarizes the results of our study.

\section{The deviating dust composition and spectral energy distribution of HD\,100546}
\label{sect:hd-energy}

The spectral appearance and dust properties of HD\,100546 differ from 
that of other HAEBE stars. We will start out with a summary of these
differences:

First, the ISO-spectra of HD\,100546 show strong and pronounced
emission, identified with the crystalline silicate forsterite,
which in abundance must be far in excess of that seen in other
HAEBE systems \citep{hd100546malfait,HerbigOverview}.

Second, \citet{processing} identified an emission component
at 8.6\,$\mu$m in the spectra of a large sample of HAEBE stars
which they attribute to silica (SiO$_{2}$), and which seems to 
be correlated with forsterite. This correlation is interpreted 
as evidence that the crystalline silicates are formed by thermal 
annealing. However, the dust properties in HD\,100546 -- and in
this respect also that of comet Hale-Bopp -- does not comply with
this correlation between silica and forsterite. This suggests that
a different formation process is responsible for the large
mass fraction of forsterite seen in both objects.

Third, from a detailed analysis of the 10\,$\mu$m spectral
region \citep{processing}, it was shown that the typical
grain size of the silicate particles emitting at these wavelengths
is larger compared to the sizes found in other HAEBE systems.
This suggests the disk in HD\,100546 is more evolved.

\begin{figure}[t!]
\centerline{\resizebox{\hsize}{!}{\includegraphics[angle=0]{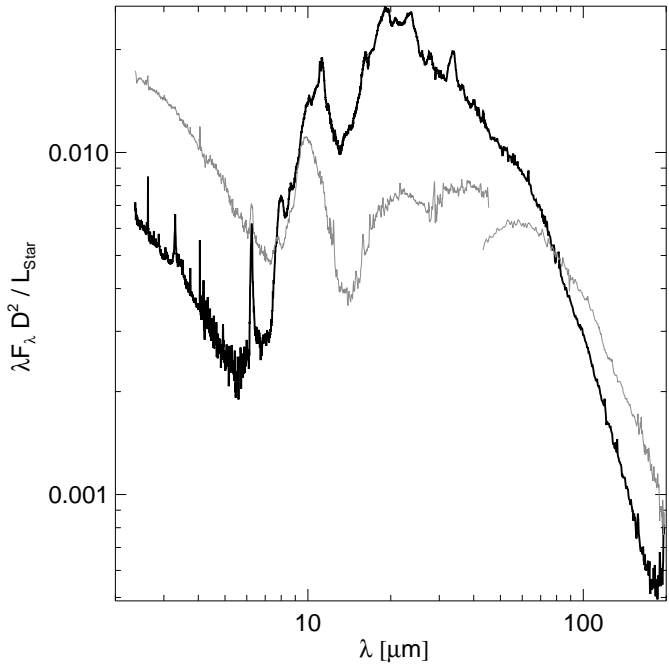}}}
\caption{ 
         Comparison between the energy distribution of AB\,Aur (light line) 
         and HD\,100546 (dark line). The spectra have been normalized to the respective
         stellar luminosities. Plotted are the combined ISO-SWS and LWS spectra. 
         Note the large difference in the near- and mid-IR luminosity of
         both systems.
        }
\label{fig-compare}
\end{figure}

Fourth, the spectral energy distribution of HD\,100546 shows
remarkable differences when compared to other HAEBE stars.
To illustrate the difference, we have plotted in Fig.~\ref{fig-compare}
the SEDs of AB\,Aurigae (light line) and HD\,100546 (dark line)
normalized to the stellar luminosity. Both dust disks intercept
about 50\,\% of the stellar light (see Table\,\ref{tbl-4}). It is 
easily seen that AB\,Aur re-emits a larger fraction of this 
intercepted light in the near-IR, while HD\,100546 shows a larger 
emission at mid-IR wavelengths. The far-IR and sub-millimetre 
luminosities are comparable. The difference in the way radiation
is redistributed by the dust essentially implies that the
stellar flux is absorbed at different locations in the circumstellar
disk. Dust grains dominating the near-IR emission have temperatures
of at least $\sim$ 1000 K. For a grain to reach such a temperature
it has to be within $\sim$ 1 AU of the central star. In the case
of AB\,Aur it is within this region that about half of the intercepted
light is reprocessed (see Table\,\ref{tbl-4}). HD\,100546 emits
a relatively modest amount of radiation in the near-IR,
however, 70\,\% of the total dust luminosity comes out in the mid-IR.
Grains emitting at these wavelengths have typical temperatures of 
$\sim$ 200 K. Assuming these grains receive direct stellar 
light this implies that the re-emitted mid-IR flux originates from 
about 10 AU from the central star.  
 
In estimating the above emitting regions, we have assumed that the
dust medium is optically thin. Let us, for the moment, hold on to this
assumption. For $\tau \ll 1$ the emission scales with dust mass,
implying that the disk of HD\,100546 is less massive in the inner
parts compared to AB\,Aur. Given that the system is older
\citep{mario1998}, it suggests that a larger fraction of the
circumstellar matter has been removed from the disk (evaporated or
accreted) and/or is incorporated into larger bodies, substantially reducing the
near-IR emission.  In an optically thick view, the emission scales not
with the dust mass but with emitting surface.  \citet{dullemond2001}
show that the near-IR excess may be due to a puffed-up inner
rim surface.  The height of this inner rim is a function of the
stellar luminosity and of the surface density in the disk near the
evaporation temperature of the dust.  The low near-IR excess of
HD100546 compared to AB Aur is then the combined effect of the smaller
stellar luminosity, and a severe depletion of the inner disk.

\begin{table}
\begin{center}
\begin{tabular}{@{} lcccc @{}}
\hline\noalign{\smallskip}
\hline\noalign{\bigskip}
&L$_{\mathrm{d}}$/L$_\star$&L$_{\mathrm{NIR}}$/L$_{\mathrm{d}}$&L$_{\mathrm{MIR}}$/L$_{\mathrm{d}}$&L$_{\mathrm{FIR}}$/L$_{\mathrm{d}}$\\
\noalign{\smallskip}
AB\,Aur&0.46&0.51&0.34&0.15\\
HD\,100546&0.51&0.17&0.70&0.13\\
\noalign{\smallskip}
\hline\noalign{\smallskip}
\end{tabular}
\end{center}
\vspace{-0.5cm}
\caption[]{Luminosity of the circumstellar dust $L_{\rm d}$ in AB\,Aur and H\,100546
           as a fraction of the stellar luminosity (first column).
           The remaining columns list fractional dust luminosities in the 
           near-IR ($\lambda<8 \mu$m), the mid-IR ($8<\lambda<50 \mu$m) and 
           far-IR ($\lambda>50 \mu$m), respectively.
          }
\label{tbl-4}
\end{table}

Concerning the mid-IR fluxes, similar considerations apply. 
\citet{dullemond2001} show that the SED of AB\,Aur can be
reproduced with a passive reprocessing flaring disk in hydrostatic
equilibrium. As the mid-IR spectrum of this star is similar
to that of many other HAEBE stars \citep{HerbigOverview}, most likely 
this model has generic value. However, the model can not explain the 
SED of HD\,100546. For a uniform flaring disk model to reproduce
the strong mid-IR emission, the opening angle of the disk has to
increase faster with radius in order to intercept more stellar
light relatively close to the star. Given the comparable sizes
of the systems \citep{augereau2001,mannings1997},
such a solution would lead to a much larger total covering fraction.
However, these are about 50\% in both systems.

A possible solution for this problem is to induce a sizable gap in the
circumstellar disk. Such a gap would produce a large rim at the far 
side of the gap (when viewed from the direction of the central star),
of which the largest part would be irradiated directly by the star
(see Fig.~\ref{fig-disk}). As a result, a larger absorbing surface is 
created and consequently the larger emitting surface that is required.
\citet{dullemond2001} show that in such a geometry indeed the
disk scale height is increased substantially, increasing the emission
from the region where the dust is $\sim$ 200 K. In Sect.~\ref{sect:disk_geometry}
we will speculate that a massive planet, which has formed in the
disk at $\sim$ 10 AU, may have created such a gap.

We conclude that HD\,100546 differs considerably from other HAEBE
systems in both dust composition and disk geometry. This implies 
that processes different from those observed in other HAEBE systems
currently dominate the HD\,100546 system. In this paper we will
try to trace these processes.

\section{Modelling}
\label{sect:hd_modelling}

The diagnostic method adopted is identical for the Herbig Be star HD\,100546
and comet Hale-Bopp. We use the radiative transfer program {\sc modust}
to model the circumstellar dust using the mode in which the material
is assumed to be optically thin. This is correct for comet Hale-Bopp,
but for the circumstellar disk of HD\,100546 this may not be
valid. However, in discussing dust properties we will focus on the
mass over temperature distribution of the material responsible for the
infrared emission. This diagnostic does not depend on the optical depth
properties of the medium and therefore the simple approximation used
here is still meaningful. For both objects the material is distributed
in a spherical shell, which for Hale-Bopp is 
positioned at the distance from the sun corresponding to that of the
location of the coma and tail at the time of observation. For recent
applications and for descriptions of techniques used in {\sc modust}, see
e.g. \citet{abaur} and \citet{kemper2001}.

The grains are irradiated by the central star, for which we use
\citet{kurucz1993} energy distributions. The Hipparcos distance to the 
B9Vne star HD\,100546 is $103 \pm 7$~pc, yielding a luminosity 
$L = 32~\lsun$ \citep{mario1998}. Comparison with evolutionary tracks,
using $\teff = 10\,500$~K, places the star on the main sequence
having an age of $>10$~Myr and mass of 2.4~\msun. In modelling Hale-Bopp,
we use Kurucz's solar model with $\teff = 5\,777$~K.

The absorption of stellar radiation at mostly ultraviolet and optical wavelengths 
and re-emission in the infrared is consistently taken into account 
assuming the particles are in radiative equilibrium, yielding the spectral 
energy distribution. A key difference of this method compared to other
often employed approaches in modelling circumstellar and cometary spectra
is that both the spectral characteristics responsible for the distinct 
features as well as those responsible for the featureless continuum are 
modelled simultaneously. Other methods usually introduce an artificial separation 
between continuum and features, fitting the spectrum combining empirical
temperatures with laboratory extinction efficiencies
\citep[e.g.][]{wooden1999,mason2001} or with template extinctions based on
measurements of interplanetary dust particles (IDP's)
\citep[e.g.][]{wooden2000}, complicating a reliable interpretation.

In the near- and mid-IR, typically carbon and large
($\ga 10$~\micron) amorphous silicate grains are responsible for
the overall continuum, while water ice is a dominant contributor
in de far-IR and sub-millimetre range. At 10~\micron\
the molecular vibrational modes of Si-O bonds in small amorphous
silicate grains ($\la 10$~\micron) produce a distinct feature,
as do silicate crystals in the mid-IR part of the spectrum. These
latter particles, however, also produce significant flux outside
of their resonances, which one can not {\em a priori} distinguish
from continuum only contributors. Continuum subtraction methods
therefore are susceptible to (systematic) errors. This may
significantly affect temperatures, abundances, and sizes derived
for especially the crystalline material. Our method does not suffer from 
this problem.
In fairness, we should mention that in other respects studies using
ground based data have advantages over the interpretation of ISO
data. For instance, the mid-IR observations analysed by Wooden
al. (2000) -- which have been obtained close in time to the ISO
data -- have much smaller beams, allowing one to study gradients
in the particle properties as a function of distance to the cometary
nucleus.

\subsection{Size and shape properties of grains}

Composition, size and shape properties of grains in proto-planetary
disks may provide important constraints on the formation history
of circumstellar dust. 

We first focus on shape properties. Very little is known about the
structure of grains in circumstellar disks. Particles may be compact
or ``fluffy'' and may be chemically homogeneous or consist of a mix 
of different materials. Information may be obtained from measurements
of grains sublimating from solar system comets nearing perihelion.
From a comparison of measurements of the angle dependence of the
scattering albedo of several bright comets with theoretical predictions
\citet{gehrz1992} found the coma grains to be consistent with
``fluffy'' aggregates of smaller compact particles.
Properties of the aggregate, such as temperature
and spectroscopic signature, will be affected by the actual degree
of fluffiness of the grain, often expressed in terms of the porosity 
factor of the particle \citep{lisse1998,harker1999}. Large porous 
grains may heat to much higher temperatures than compact grains of the 
same mass and may show spectroscopic characteristics similar to the 
smaller compact particles constituting the fluffy aggregate. Theoretical 
considerations suggest cometary grains to be extremely fluffy with
porosity factors $0.93 \la P \la 0.975$ \citep{greenberg1990},
effectively implying that some spectroscopically deduced properties 
relate to the smaller compact units making up the porous aggregate
\citep[e.g.][and references there in]{brucato1999}.
In our modelling we will therefore concentrate on (small)  compact 
particles, leaving open the possibility that they may be part of larger
fluffy aggregates.

Regarding the chemical homogeneity of the small particles, 
\citet{li1998} pointed out that {\em in situ} mass spectra of comet
P/Halley 1986 III dust showed that carbonaceous and silicate materials
were mixed on fine scales, suggesting these two species are not
physically separated. If this is the case, the components are likely in
thermal contact. The question of thermal contact is especially 
relevant with respect to the magnesium over iron content of the
silicate material, a major constituent of circumstellar dust, as 
the absorption properties
of olivine (Mg$_{2x}$Fe$_{2-2x}$SiO$_{4}$, with $x$ between 0 and 1) and pyroxene 
(Mg$_{x}$Fe$_{1-x}$SiO$_{3}$) sensitively depend on this ratio.
The pure magnesium silicates forsterite (Mg$_{2}$SiO$_{4}$)
and enstatite (MgSiO$_{3}$) are optically much more transparent than
iron rich silicates, implying that if Fe-rich and Fe-poor materials
coexist as separate particles -- with comparable shape and size -- the
Fe-rich dust will reach significantly higher temperatures. 

We have opted to treat the different chemical species as physically
separated. In the case of Hale-Bopp this allows to investigate 
whether the forsterite crystals, of which the spectroscopic 
signatures are prominently visible in the mid-IR spectrum, are
in thermal contact with the bulk amorphous silicate material, which is
likely to contain a significant fraction of iron, by determining
the mass-averaged temperatures of both species. If significantly
different, one may assume the components to be physically
separated; if similar temperatures are found, it is likely they
are in thermal contact.

For each particle, we calculate the extinction properties from 
optical constants determined in laboratory experiments, as listed in Table~\ref{tbl-1}. 
We assume spherical grains, for which we use Mie calculations, or a continuous 
distribution of ellipsoidal grains, for which we use CDE calculations,
to determine the absorption and scattering coefficients 
(see \cite{bohren1983} for al full review on these methods).

As we treat the CDE particles in the Rayleigh limit, we can not study 
grain size effects.
For spherical particles this assumption is not required.
The spherical particles of our multi-component mixture of grains
range between minimum size \amin\ and maximum size \amax\
and are distributed following a power-law, i.e.
\begin{equation}
   n(a) \propto \left( \frac{a}{a_{\rm min}} \right)^{-m}.
\end{equation}
Theoretical calculations predict this type of size distribution
whenever there is shattering and coagulation of grains through
grain-grain collisions (Biermann \& Hartwit 1980). To get some feeling
for the value of the power-law index, extinction observations imply 
a size distribution with $m$ = 3.5 for interstellar grains 
\citep{MRN}. 

\subsection{Chemical composition of grains}

The grain species used to model the spectra of both objects are
listed in Table~\ref{tbl-1}. The dust composition is
similar to that used by \citet{abaur} to model the isolated
Herbig Ae stars AB\,Aurigae and HD\,163296, though more appropriate sources
for the optical constants of crystalline silicates and water ice are 
used. Of the two isolated Herbig stars mentioned only in HD\,163296 
a small amount of forsterite could be identified, modeling of which 
did not sensitively depend on the adopted optical constants. However, 
for the two objects investigated here -- which show prominent features 
of olivine -- differences in laboratory measurements are important.
To model the crystalline silicate features we tried three sets of data:
\citet{servoin1973} measured the optical properties of forsterite (Mg$_2$SiO$_4$; Fo100).
\citet{steyer1974} used a natural olivine sample, which was estimated to contain a 
small amount of iron, i.e. $x = 0.91$ (Fo91). The sample used by \citet{mukai1990}
contains an almost equal amount of iron ($x = 0.90$; Fo90). These measurements only cover the
IR wavelengths.  To estimate the optical properties of
the grains at visual and UV wavelengths, where most of the stellar light is absorbed,
we used measurements of comparable materials from several other sources, referenced in
Table~\ref{tbl-1}.

Neither in AB\,Aur nor in HD\,163296 the 44 \micron\ feature 
characteristic for crystalline H$_{2}$O ice was found, prompting the
use of amorphous water ice in modeling their circumstellar environment. 
In both HD\,100546 and Hale Bopp, however, the 44 \micron\ feature 
appears to be visible, making it more appropriate to assume the water 
ice in these sources to be crystalline.

\begin{table}
\begin{center}
\caption[]{Overview of dust species used. For each component we specify
           its latice structure, the
           wavelength interval over which optical constants are
           measured, and the destruction temperature. 
\label{tbl-1}
          }
\bigskip
\begin{tabular}{@{} llcrl @{}}
\hline\noalign{\smallskip}
\hline\noalign{\bigskip}
Species                & Latice    & Wavelength & T$_{\rm destr}$ & Ref.  \\
                       & Structure & [\micron]  & [K]            &       \\
\noalign{\smallskip}
\hline\noalign{\smallskip}
[Mg,Fe]SiO$_{4}$          & A     & 0.2--500   & 1100           & (1)   \\
FeO                    & C     & 0.2--500   & 1000           & (2)   \\
C                      & A     & 0.1--800   & 1000           & (3)   \\
H$_{2}$O               & C\,(ice)& 0.05--10$^{3}$ & 150& (4) \\ 
&&&&\\
Forsterite             & C     & 0.04--3   & 1400    & (5)   \\
                       &       & 3--250     &                & (6)   \\
&&&&\\
Olivine (Fo91)         & C     & 0.04--3 & 1400      & (5)   \\
                       &       & 3--250     &                & (7)   \\
&&&&\\
Olivine (Fo90)         & C     & 0.01--0.3 & 1400    & (8)   \\
                       &       & 0.3--2     &                & (9,10)\\
                       &       & 7--200     &                & (11)  \\
&&&&\\
Fe                     & M     & 0.1--10$^{5}$ & 1500        & (12)  \\
\noalign{\smallskip}
\hline
\end{tabular}
\end{center}
\footnotesize
 Abbreviations used to designate the latice structure: 
 A = Amorphous; C = Crystalline; M = Metallic.

\smallskip
\noindent
{\em References:}
 (1) \citet{dorschner1995};
 (2) \citet{henning1995};
 (3) \citet{preibisch1993};
 (4) \citet{bertie1969};
 (5) \citet{scott1996};
 (6) \citet{servoin1973};
 (7) \citet{steyer1974};
 (8) \citet{huffman1973};
 (9) \citet{jones1976};
 (10) \citet{rogers1983}
 (11) \citet{mukai1990};
 (12) \citet{henning1996}.
\end{table}

\section{Results}
\label{sect:hd-results}


\subsection{HD\,100546}

\begin{figure}[t!]
\centerline{\resizebox{\hsize}{!}{\includegraphics[angle=0]{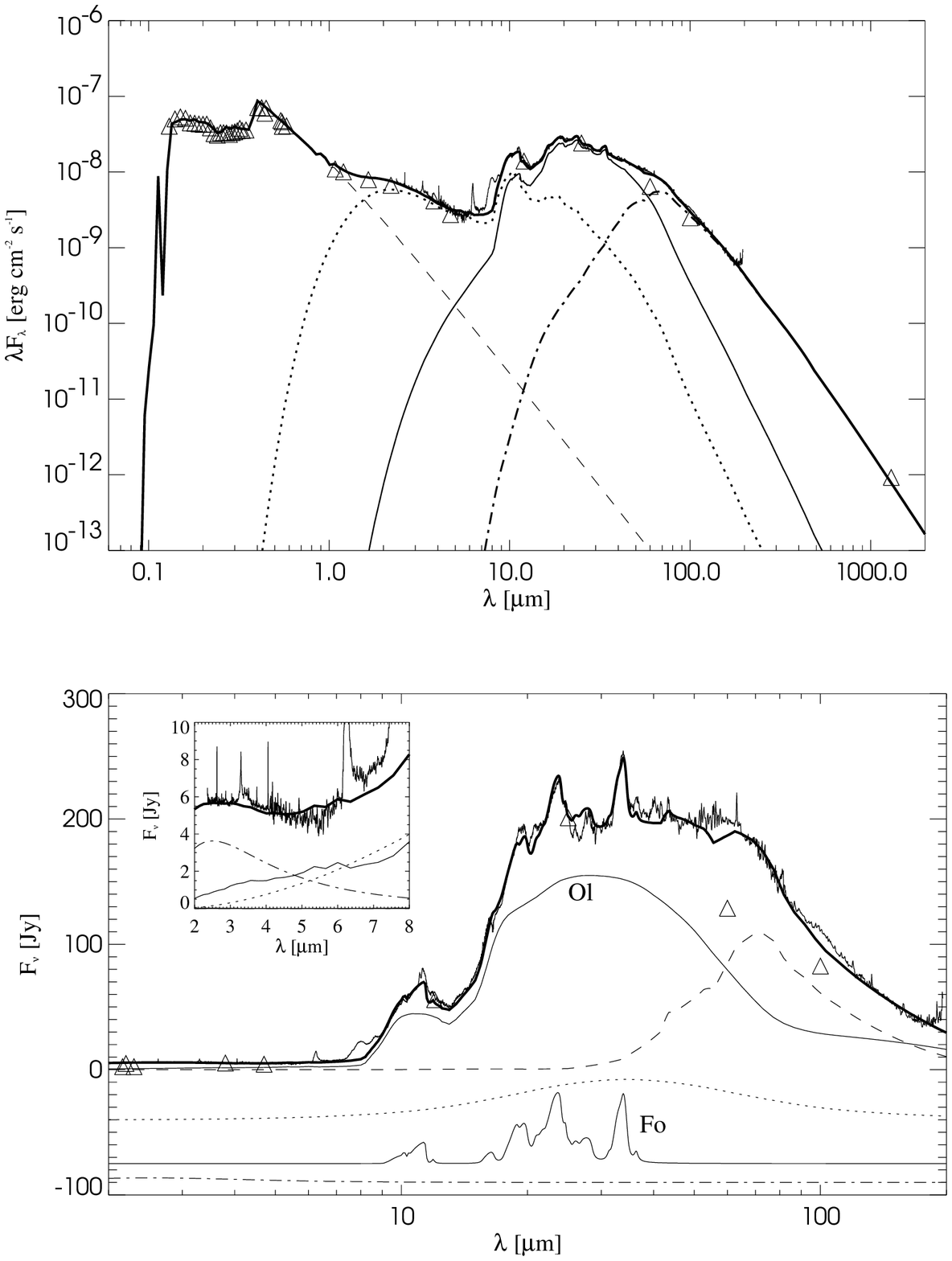}}}
\caption{The top panel shows our best model fit to the spectral energy 
distribution of HD100546 (thick solid line). 
Indicated with the dashed line is the spectrum of the central star. 
Triangles indicate ground-based and IRAS photometry. Also plotted are the 
the ISO-SWS and LWS spectra. The dashed line, the thin solid line and 
the dashed-dotted line represent the contributions of the small grain 
components in zone~1, zone~2 and the large grain component, respectively. 
The bottom panel shows our fit to 
the ISO spectra in detail. 
Also plotted in this figure are the relative contributions of the individual 
dust components. Indicated with a solid line are the contributions 
of amorphous olivine and forsterite marked in the figure 
with Ol and Fo, respectively. The contribution of water ice, carbonaceous grains and metallic 
iron are represented by the dashed, dotted and dashed-dotted lines, respectively.
The curves of the carbonaceous, forsterite and metallic iron grains are offset
by -40, -75 and -90 Jy, respectively.  The inset shows the 2 to 8~\micron\ region.
} 
\label{fig-spectra}
\end{figure}

Plotted in Fig.~\ref{fig-spectra} is our best model fit to the SED
of HD\,100546. The resulting model parameters are listed in Table~\ref{tbl-2}. The top panel 
shows the entire SED. A dashed line represents
the Kurucz model for the central star.
The lower panel shows the ISO-SWS and LWS wavelength region. Indicated
are the contributions to the spectrum of the individual dust components as listed in 
Table~\ref{tbl-2}. To fit the SED a bi-modal grain size distribution is required, similar
as to that found in previous analysis of HAEBE stars \citep{abaur,thesis}. The small ($<10$~\micron) grains
dominate the SED shortward of $\sim 40$\micron, while the large (up to $200$~\micron) grains
dominate at the longest wavelengths.

\begin{table*}
\begin{center}
\begin{tabular}{@{} ccccccccccc @{}}
\hline\noalign{\smallskip}
\hline\noalign{\bigskip}
Component:&\multicolumn{7}{c}{Small grains}&~~~~~~&\multicolumn{2}{c}{Large grains}\\
\hline\noalign{\smallskip}
&\multicolumn{6}{c}{two component density structure}&&&\\
&\multicolumn{3}{c}{zone 1}&\multicolumn{3}{c}{zone 2}&\\
${\rm M}_{\rm dust}/\msun$&\multicolumn{3}{c}{$5.1\cdot 10^{-9}$}&\multicolumn{3}{c}{$6.7\cdot 10^{-7}$}&&&\multicolumn{2}{c}{$6.5\cdot 10^{-5}$}\\
Dust species&R [AU]&$\rho_0$ [gr cm$^{-3}$]&${\rm M}_{\rm frac}$&R [AU]&$\rho_0$ [gr cm$^{-3}$]&${\rm M}_{\rm frac}$&a [$\mu$m]&&${\rm M}_{\rm frac}$&a [$\mu$m]\\
&\multicolumn{3}{c}{------------------------------------}&\multicolumn{3}{c}{------------------------------------}&&&&\\
Amorph. silicate&0.3--9.8&$1.6\cdot 10^{-17}$&0.88&9.8--43&$2.7\cdot 10^{-18}$&0.7&0.01--10&&0.53&10--200\\
Carbon&0.3--9.8&$1.7\cdot 10^{-18}$&0.09&9.8--43&$3.1\cdot 10^{-19}$&0.08&0.01--10&&0.05&10--200\\
Metallic iron&0.3--4.9&$4.2\cdot 10^{-19}$&0.01&--&--&--&0.01--0.1&&--&--\\
Forsterite&0.3--24&$6.0\cdot 10^{-21}$&0.02&24--600&$1.4\cdot 10^{-21}$&0.19&--&&--&--\\
Water ice&--&--&--&9.8--43&$1.15\cdot 10^{-19}$&0.03&0.01--10&&0.42&10--25\\
\noalign{\smallskip}
\hline
\end{tabular}
\end{center}
\caption[]{Best fit model parameters of HD\,100546. Listed are the parameters 
defining the density and grain size distribution, the chemical composition and the mass 
fraction ${\rm M}_{\rm frac}$ of the individual dust species. Both the small grain and 
large grain component have a power law density distribution $\rho(r)\propto r^{-1}$
and grain size distribution $n(a)\propto a^{-2}$. The radial extent of the small grain 
component is given in the table, large grains are present between 28--380~AU with a density 
at the inner boundary of $\rho_0 = 1.5\cdot 10^{-18}$. Note that the densities given at 
the inner radius $\rho_0 = \rho(\mathrm{R_{in}})$ of zone 1 assume all dust species to be
present. However for temperatures greater than the destruction temperature $T_{\mathrm{destr}}$ 
(see Table~\ref{tbl-1}) the actual densities are slightly lower.
 \label{tbl-2}}
\end{table*}

\subsubsection{The small grain component}
\label{subsec:hd_small_grains}

Though the small grain component contains only a minor fraction of the total dust mass
($\sim1\%$) it dominates the SED at near- and mid-IR wavelengths. To reproduce the 
observed fluxes
in this wavelength range, a bi-model density structure is required.
As can be seen from Table~\ref{tbl-2}, the small grains are distributed between 0.3--9.8~AU,
and between 9.8--43~AU, from here on referred to as zone 1 and zone 2, respectively.
The contributions to the SED of zone~1 and 2 are plotted in Fig.~\ref{fig-spectra}.
The inner boundary of zone 1 at 0.3\,AU  is determined by the grain destruction temperatures
of the individual dust species as listed in Table~\ref{tbl-1}. A similar dust component as that in zone~1
is also present in AB\,Aur, HD\,163296 and HD\,104247 \citep{abaur,thesis}. The sudden increase
in density at 9.8~AU, marking the onset of zone~2, is found only in HD\,100546. It is
this additional component that produces the much larger mid-IR luminosity compared to the other
HAEBE systems as discussed in Sect.~\ref{sect:hd-energy}. 
As can be seen from Fig.~\ref{fig-spectra}, the emission from zone~2 completely
dominates the SED between $\sim$20 to 50~$\mu$m.
The resulting total mass
of $6.8\cdot 10^{-7}$~\msun\ in small grains is two to three orders of magnitude larger
than found in the studies mentioned above. 

A major fraction of the dust emission seen in HAEBE systems
is due to amorphous silicates. Though it is difficult to determine the exact nature of the 
amorphous material, an excellent fit can be made if we use the optical properties of a silicate
glass with an olivine stoichiometry (see also Table~\ref{tbl-1}). The grain size of the 
amorphous silicate is well constrained by the shape and strength of the 10~\micron\ 
silicate feature, and the flux ratio between the 10 and 18~\micron\ bands. 
Apart from the amorphous silicates an additional source of continuum emission
is required. We added this in the form of carbonaceous grains. Since this material has no
distinct spectral features, the grain size and consequently, mass fraction, are less well
established and are uncertain within a factor of two. In our best fit model we assumed 
an equal grain size distribution as for the amorphous silicate. 
The band at 43.3~\micron\ in the spectrum of HD\,100546 indicates the presence of a population
of small ice grains. Given the ice sublimation/evaporation temperature of 150~K, 
water ice can only be present in zone~2 outside of the ``snow limit''.
For the small ice grains we also assumed the same grain size distribution as for
the amorphous material. Interesting to note is that \citet{smith1994} showed that the
ice band can only shift to the observed wavelength of 43.3~\micron\ 
if amorphous ice grains where heated to 
the crystallization temperature and then cooled to $\sim$50~K. One could interpret this
as a local (i.e. within the ``snow'' regime) heating event or local radial mixing.

As one can see from the inset
in Fig.~\ref{fig-spectra}, the near-IR flux shortward of 4~\micron\ is dominated by 
the emission
from metallic iron grains. These grains are the only refractory dust species  which are
stable at the required high temperatures ($\sim 1500$~K), and that have sufficient opacity 
at near-IR wavelengths to emit in this spectral region. We had to constrain the spatial 
distribution of the metallic iron grains to within 4.9~AU. 
The grain temperature of the iron 
grains is between 1500 and 750~K. This suggests that metallic iron is formed at high
temperatures, most likely by solid state reduction of iron bearing silicates heated above
the glass temperature at the very inner parts of the circumstellar disk.

\subsubsection{The crystalline silicate component}
\label{sect:hd_small_Xsil}

\begin{figure}[t!]
\centerline{\resizebox{\hsize}{!}{\includegraphics[angle=0]{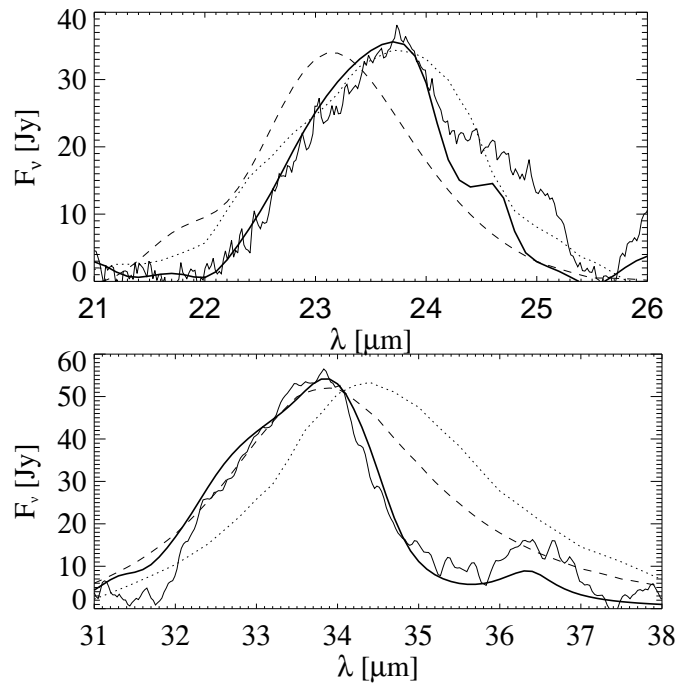}}}
\caption{The 24 and 34~\micron\ forsterite bands of HD\,100546. Overplotted are
the best fit silicate bands using the data of \citet{servoin1973} (solid line),
\citet{steyer1974} (dotted line) and \citet{mukai1990} (dashed line). In the latter
model spherical grains where assumed with a size of 0.01\,\micron. The other models have
a CDE shape distribution.
}
\label{fig-bands}
\end{figure}

In the sample of isolated HAEBE stars presented in \citet{HerbigOverview}, HD\,100546
shows the most prominent features of crystalline silicates. This enables one, at least in principle,
to determine the exact chemical composition and crystalline structure of the silicates.
We used three sets of laboratory data listed in Table~\ref{tbl-1}, to determine the best fit
to the crystalline silicate bands. Plotted in Fig.~\ref{fig-bands} are the 24 and 34~\micron\
silicate bands of HD\,100546, compared with our best fit models for the three data sets.
As one can clearly see, the model based on the measurements of \citet{servoin1973}
provides the best match with the observations. For this model and the model using the
\citet{steyer1974} data set, the CDE shape distribution 
(see Sect.~\ref{sect:hd_modelling})
gives a much better fit than do spherical grains. In the model using 
the \citet{mukai1990} data, MIE calculations produce the best
match. We conclude that the crystalline silicate features seen in the ISO spectra
of HD\,100546 are due to non-spherical forsterite grains. In our best fit, presented in
Fig.~\ref{fig-spectra} and Table~\ref{tbl-2}, we used this latter model for the crystalline silicate.

The forsterite grains also require a bi-modal density structure (zone~1 and zone~2).
As can be seen from Table~\ref{tbl-2} the crystalline mass fraction is much higher
in zone~2 than in zone~1, indicating that the abundance of crystalline material is increasing 
outwards.
In our best fit model the forsterite grains are not co-spatial with the other dust species. 
The spatial distribution
of forsterite grains has to extent out to 600~AU, well beyond the observed disk limit of
380~AU \citep{augereau2001}. This, however, could be an artifact of the grain 
model we used for calculating the optical properties of the forsterite grains. 
The CDE approximation assumes the grains to be in the Rayleigh
limit, and consequently the optical properties of the particles do not depend on grain size. 
However, outside the
Rayleigh limit, particle size has a strong effect on grain temperature. Larger grains have to
be placed much closer to the central star than smaller grains to reach the same equilibrium
temperatures. 
The grains can however not be much larger than 1~\micron\ to be able to reproduce the
observed spectral features. This is smaller than the typical grain size of the amorphous 
material. Our best estimate for the effect of grain size on the spatial distribution, 
using the measurements of \citet{mukai1990} indicates that the crystalline grains could
be moved inwards to within the observed disk limits, but not far enough to be co-spatial 
with the amorphous silicate.

The grain size distribution also influences the mass fraction of the dust species. The model assuming
a maximum grain size of 2~\micron\ for the forsterite grains would increase the mass in
crystalline dust by a factor of five. This would imply that as much as 50~\% of the total mass 
of the small grain component could be crystalline silicates.

\subsubsection{The large grain component}
\label{subsec:hd_large_grains}

To reproduce the observed slope and fluxes at mm wavelengths a dust component with
a large grain size up to 200~\micron\ is required. 
Their contribution to the SED of HD~100546 can be seen in Fig.~\ref{fig-spectra}.
Due to this large grain size, one can 
not determine its exact composition from ISO spectroscopy, since such large grains
will radiate as black bodies at these wavelengths. We used the dust species found 
spectroscopically to model the large grain component and assumed their relative
mass ratios to be identical to that of the small grain component. The mass
fractions of the large amorphous silicate and carbonaceous grains, listed in Table~\ref{tbl-2},
are therefore uncertain.
Similarly, it is not possible to determine if these large
grains contain crystalline material as this will also result in black body emission.
We therefore did not use crystalline silicates in the large grain component. However, due to the
presence of the emission band at 60~\micron\ attributed to crystalline water ice grains, 
the mass fraction of this dust species can be determined. The maximum size the ice
grains can have and still produce a 60~\micron\ band, is 25~\micron. This is much smaller than the 
maximum size required to fit the slope of the SED at mm wavelengths.
There are however some problems with the 60~\micron\ band as we will discuss in 
Sect.\,\ref{sect:hd_residue}.

We have constrained the spatial extent of the large grains using the disk size observed
by \citet{augereau2001} with the HST/NICMOS2 instrument. To fit the observations,
a gap of 28~AU is required between the central star and the dust shell containing the large
grains. As noted in previous studies \citep{abaur}, this gap is most likely an artifact of
the assumption made in this study that the medium is optically thin in the radial direction. 
In HD\,100546 we see the disk fairly face on \citep[51$^{\circ}$][]{augereau2001}, implying
that in the line-of-sight the medium may well be optically thin. Therefore, the
derived mass over temperature distribution $T_{\rm d}(m)$ remains meaningful 
(albeit not the $T_{\rm d}(r)$ structure).
As suggested by \citet{abaur} the large grains have most likely settled
to the disk mid-plane, extending all the way to the disk inner edge. Recent full 2D
radiative transfer calculations \citep{thesis} confirm this qualitative picture.
These calculations also show that the $T_{\rm d}(r)$ of the hot small grains,
residing in the optically thin surface layer of the disk, is in agreement with the simple spherical optically thin
model adopted in this paper.

\subsubsection{The mass temperature distribution}
\label{sect:hd-discussion-mass}

\begin{figure}[t!]
\centerline{\resizebox{\hsize}{!}{\includegraphics[angle=0]{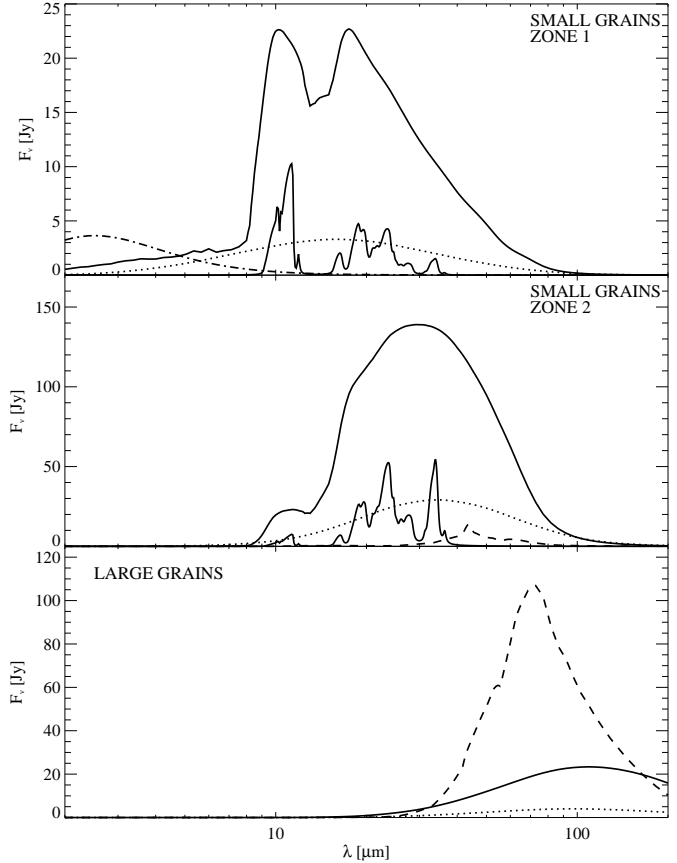}}}
\caption{Contributions to the model SED of the individual dust species per zone.
The first two panels show the contributions
of the small grain component with sizes $< 10$~\micron\ located in zone~1 and zone~2, 
respectively.  The lower panel shows the distribution of the large grains ($> 10$~\micron).
The lines represent the same materials as used in the lower panel of 
Fig.~\ref{fig-spectra}.
The top solid lines indicate the contributions of amorphous silicates 
and the lower solid lines the contribution of forsterite. 
\label{fig-spectra-individueel}
}
\end{figure}

\begin{figure*}[t!]
\resizebox{12cm}{!}{\includegraphics[angle=0]{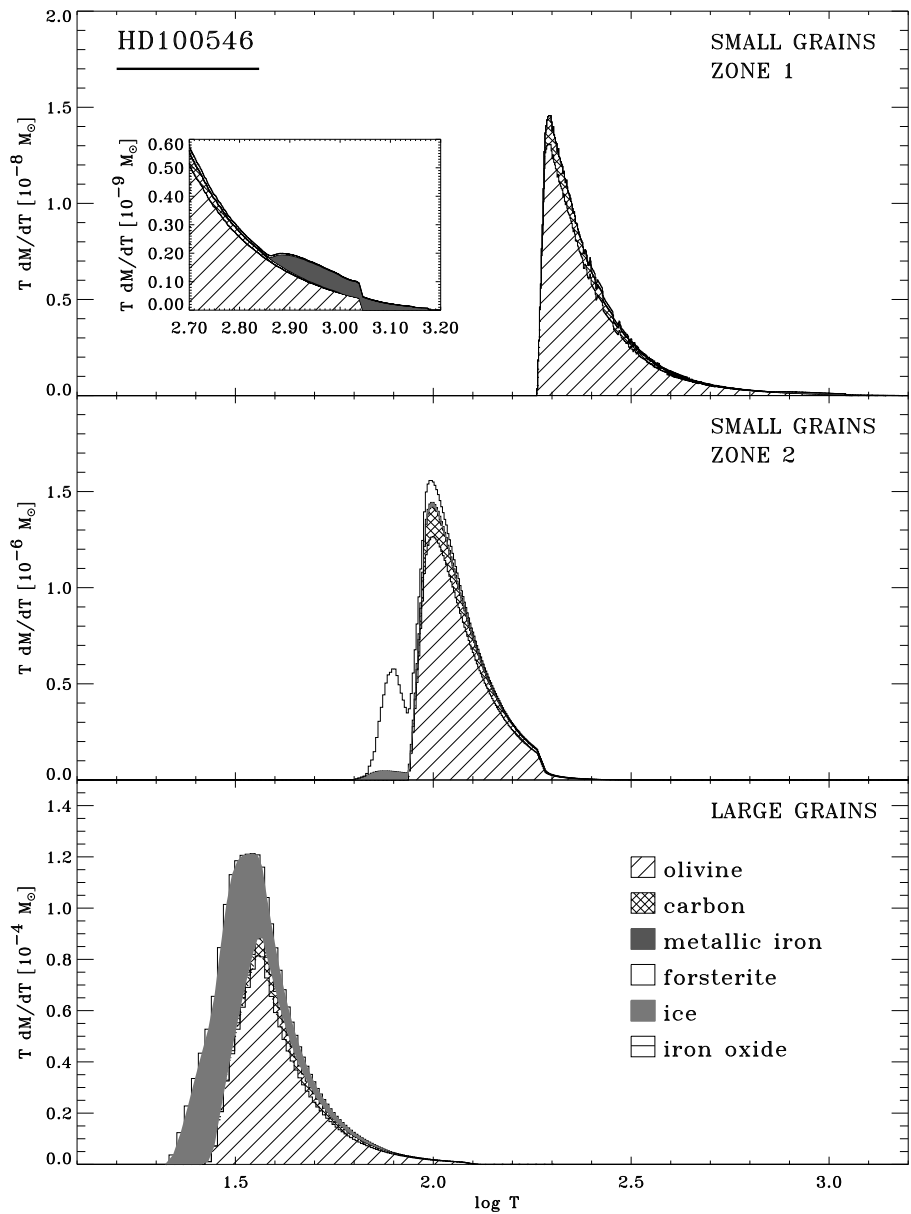}}
\hfill
\parbox[b]{55mm}{
\caption{Cumulative dust mass over temperature distribution of HD\,100546. 
Indicated in the panels are the relative contributions of the 
individual dust species. The first two panels show the contributions
of the small grain component with sizes $< 10$~\micron\ located in zone~1 and zone~2, respectively.
The lower panel shows the distribution of the large grains ($> 10$~\micron).
The inset in the top panel shows the high temperature dust component of HD\,100546.
\label{hd100546:fig-mass1}
}}
\end{figure*}

\begin{figure*}[t!]
\resizebox{12cm}{!}{\includegraphics[angle=0]{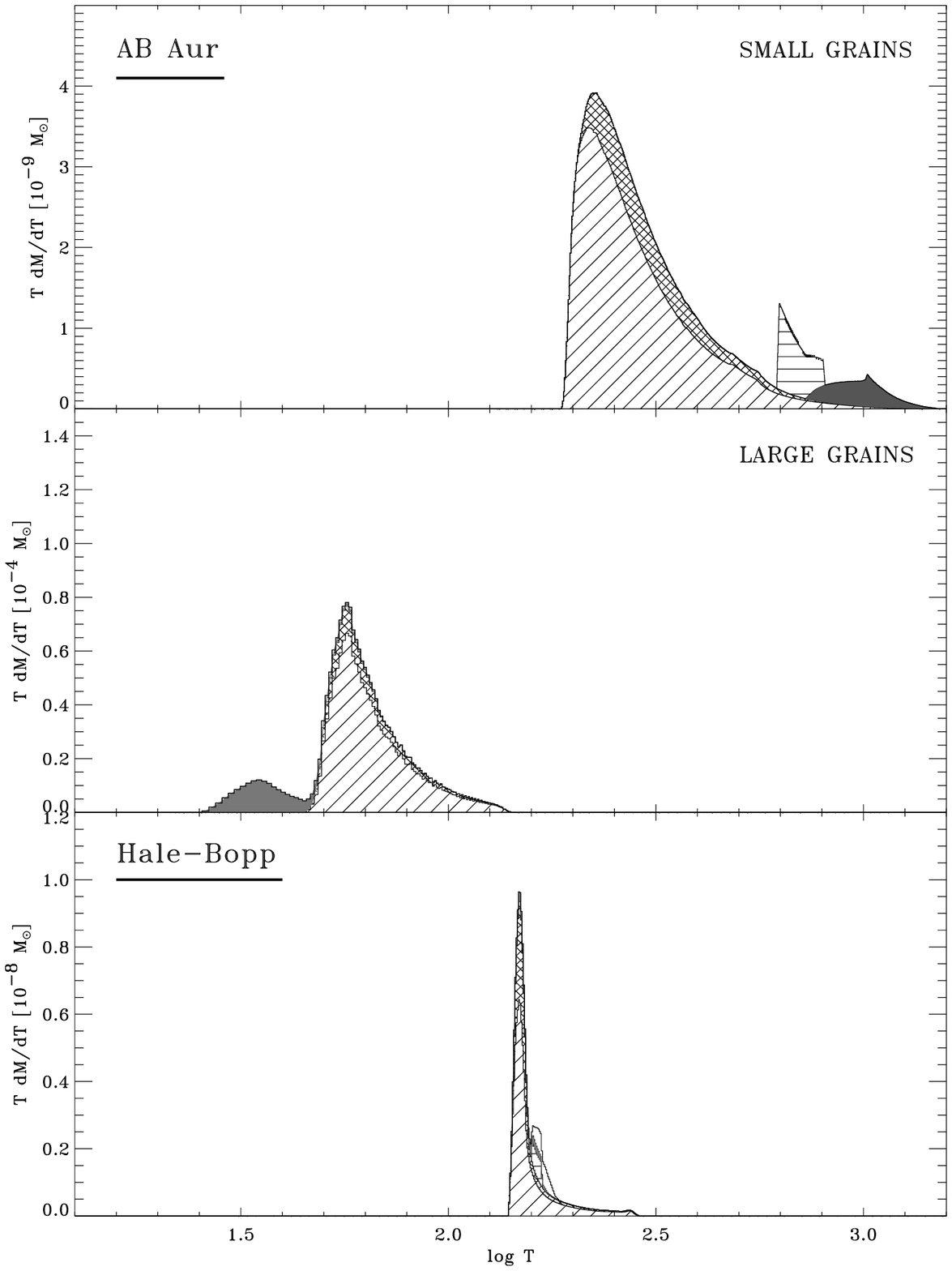}}
\hfill
\parbox[b]{55mm}{
\caption{Cumulative dust mass over temperature distribution of AB\,Aur and comet Hale-Bopp.
The two top panels show the distribution for AB\,Aur for small ($<$\,10\,\micron) and 
large ($>$\,10\,\micron) grains, respectively. 
The bottom panel shows the distribution for the solar system comet Hale-Bopp. 
Indicated in the panels are the relative contributions of the individual dust species
(see Fig.~\ref{hd100546:fig-mass1} for identification).
\label{hd100546:fig-mass2}
}}
\end{figure*}

The SED as presented in Fig.~\ref{fig-spectra} is determined by the mass-temperature distribution
of the circumstellar dust. Irrespective of the assumed model geometry, this distribution has to
be reproduced for dust that is contributing to the observed emission features.  
Plotted in Fig.~\ref{hd100546:fig-mass1} is the derived mass over temperature distribution 
of the best fit model. The upper two panels show the mass-temperature distribution of 
the small grain component, the lower
panel shows the same for the large grains.
Indicated in the figure are the individual contributions of each species.
The vertical axis unit is chosen in such a way that the integral of the cumulative mass-temperature
distribution equals the total dust mass.
We have plotted in Fig.~\ref{fig-spectra-individueel} the relative contribution to the SED
of the individual dust species to get a feeling of their importants to the overall fit.
Plotted in Fig.~\ref{hd100546:fig-mass2} for comparison is the mass-temperature 
distribution of AB\,Aur as derived by \citet{abaur}. 
The first panel shows the mass-temperature distribution of the small grains, 
the second panel that of the large grain component.

A comparison between Figs.~\ref{hd100546:fig-mass1} and \ref{hd100546:fig-mass2}
shows that the dust in HD\,100546 is characterized
by a bi-modal mass-temperature distribution, similar to that in the AB\,Aur system.  
The inset in the top panel of Fig.~\ref{hd100546:fig-mass1}
shows the mass distribution
at temperatures above 500~K, producing the near-IR emission. This component, dominated by hot iron or 
iron oxide grains, is very similar for both objects.
\citet{HerbigOverview} have shown that the near-IR spectrum of HAEBE systems is generic, suggesting a
similar geometry and/or dust composition at the very inner parts of the circumstellar dust region.
However, the total mass contained in the small grains in HD\,100546 is almost three orders of magnitude 
larger.
This excess mass is located in zone~2 (see Table~\ref{tbl-2}) of the small grain component, which is not
present in AB\,Aur and is the cause for the much larger mid-IR excess as observed in HD\,100546.

Focusing on the mass-temperature distribution of the small grains, one notices the
difference in the distribution between the amorphous dust and the forsterite grains. To fit the SED, the
crystalline silicate must have a much lower temperature than the amorphous grains.
This result holds irrespective
of the adopted laboratory measurements of the crystalline silicate. This mass-temperature distribution
imposes the small forsterite grains to have a larger radial extent compared to the amorphous dust.

The large grain component of HD\,100546 has a distribution similar to that of AB\,Aur. The total mass
contained in this component is the same within a factor of two. The small difference in dust mass
between both systems is due to a grain size difference, where in AB\,Aur the maximum grain size
is smaller by a factor of two.

\subsubsection{The residual spectrum}
\label{sect:hd_residue}

\begin{figure}[t!]
\centerline{\resizebox{\hsize}{!}{\includegraphics[angle=0]{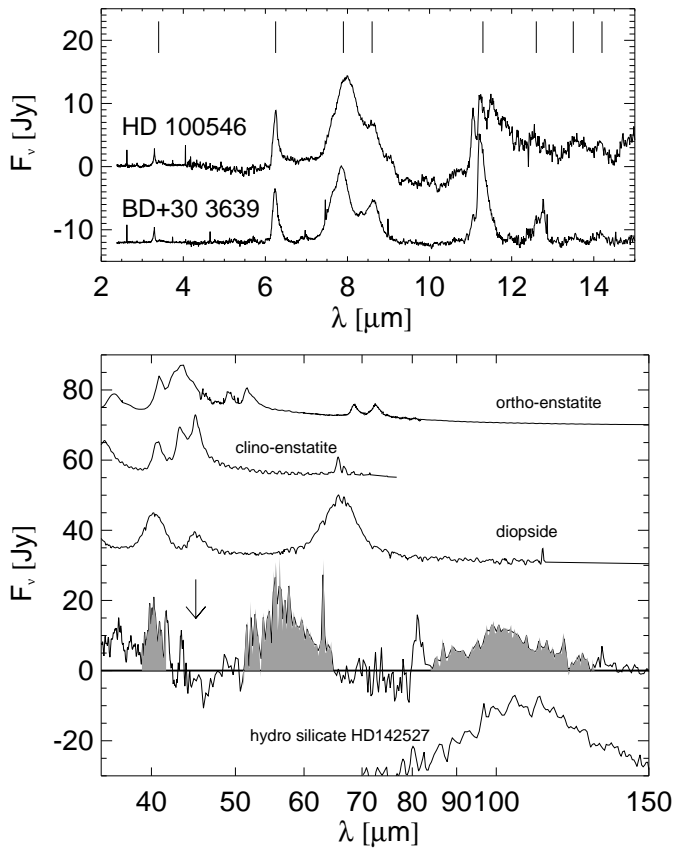}}}
\caption{Residual of the ISO-SWS and LWS spectrum of HD\,100546 after subtraction
of our best fit model.
The top panel shows the 2 to 15~\micron\ region clearly
showing the PAH emission bands (indicated with vertical lines).
For comparison we also plotted the PAH spectrum of BD+30~3639 (offset by -12 Jy).
The bottom panel shows the 35 to 150~\micron\ region. Indicated with an
arrow is the connection between the ISO-SWS and LWS spectra. Also plotted
in the figure are the absorption spectra of several silicates of interest
multiplied by a 70 K black body, and the emission band in the spectrum of HD\,142527 identified
with hydro silicates \citep{malfait1999}. The shaded areas indicate features in the residual spectrum,
discussed in Sect.~\ref{sect:hd_residue}.
}
\label{fig-residue}
\end{figure}

Plotted in Fig.~\ref{fig-residue} is the residual spectrum of HD\,100546, after subtraction
of the best fit model to the ISO-SWS and LWS spectra. Shown in the top panel is the 2 to 15~\micron\
region in which the emission bands from polycyclic aromatic hydrocarbons 
(PAHs; marked with the vertical lines) are clearly visible. For comparison we also plotted the PAH spectrum
of the [WC] star BD+30~3639, which may serve as a template of a generic PAH spectrum. Apart from
the bands at 6.25, 7.9 and 11.3~\micron, also PAH features at longer wavelengths can be 
observed \citep[see also][]{hony2001}.

The lower panel shows the 35 to 150~\micron\ region. Marked with  shaded areas
are residual emission bands. The band at $\sim$100~\micron\ has been interpreted as evidence for
the presence of hydro silicates \citep{hd100546malfait}. If correct, this can be seen as evidence for the
presence of larger bodies on which the hydro silicates are believed to have formed by aqueous alteration
of liquid water. For comparison we plotted the solid-state band of HD\,142527 which is identified with 
hydro silicates \citep{malfait1999}. As one can see the width and position are similar for both
bands.
However, a secure identification of this rather broad feature strongly depends on the exact
run of the underlying continuum. A small change in the location of this continuum may change both
the strength and width of the feature or may even cause it to disappear.

The residual at $\sim$55~\micron\ could indicate the presence of an emission band of a 
species not treated in our models. It is, however, more likely that this residual 
originates from the poor fits of the
ice bands at 43 and 60~\micron. Though the positions of the observed ice features 
coincide well with our ice model,
the width and relative strength of the 43 and 60~\micron\ bands do not. This could reflect limitations
of our simple ice model, e.g. the assumptions of homogeneous spherical grains. 
Also the temperature of the material at which the optical properties where 
measured (100~K for our data) will influence the strength and width of the features. The observed
position of the ice band at 43.3~\micron\ indicates that the 
ice is $\sim$50~K \citep{smith1994}.
An additional problem is that the ice features appear in two separate instruments, 
LWS and SWS. These instruments connect at 45\,\micron (indicated with an arrow in 
Fig.~\ref{fig-residue}). The error in the absolute flux calibration between both 
instruments is 10--15~\%. Lowering the LWS spectrum with this percentage
would enable us to fit the ice bands without the residue at $\sim$55~\micron. 
Interesting to note is that the LWS spectrum would then also agree better with the IRAS 
photometry which is now below the LWS spectrum, as can be seen 
from Fig.~\ref{fig-spectra}. However, we stress that from the ISO observations there is no strong
evidence for a jump between the ISO-SWS and LWS spectra at $\sim$45~\micron.

A third feature in our residual spectrum appears at 40.3~\micron. 
This band could be due to crystalline dust species other than forsterite. 
Plotted in Fig.~\ref{fig-residue} are the spectra of diopside and
ortho- and clino-enstatite \citep{koike2000}. All these species show bands in the 40~\micron\ region.
The 40~\micron\ band of diopside coincides nicely with the residual spectrum 
but would also produce a band at 65~\micron\ at the location of the water ice band, 
making the problem at 55~\micron\ even more severe.
If one of the suggested materials is responsible for the 40~\micron\ band, 
it has to be cold, i.e. $\la$\,70~K, as at shorter wavelengths
no evidence for crystalline silicates other than forsterite can be found. 
Finally we mention that the bands 
at 40.3 and 43.3~\micron, coinciding with the ice band, may be due to forsterite. 
This suggestion is based on  mass absorption spectra taken by \citet{jaeger1998},
which show evidence for bands at these wavelengths. Though these features are weak,
given the overall strength of the forsterite bands this could be a possibility.

\subsection{Hale-Bopp}
\label{sect:hd_HB}

\begin{figure}[t!]
\centerline{\resizebox{\hsize}{!}{\includegraphics[angle=0]{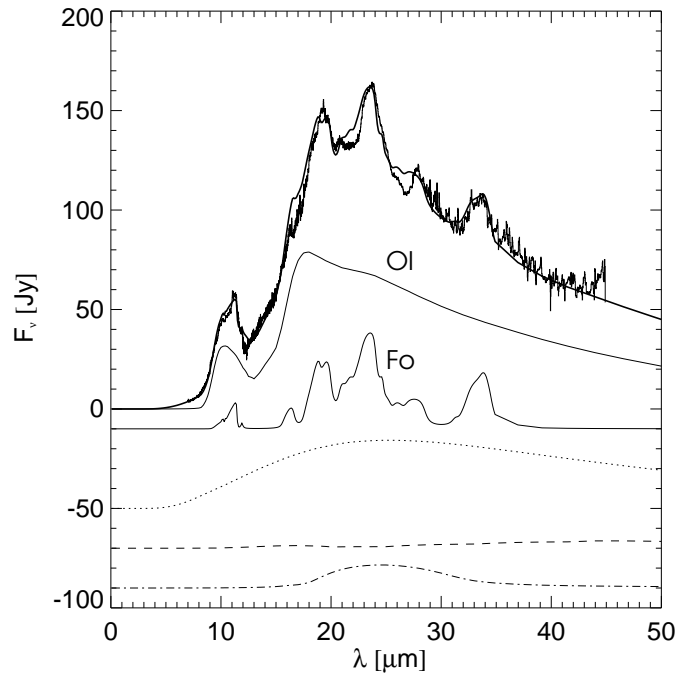}}}
\caption{Best model fit to the ISO-SWS spectrum of Hale-Bopp. Also plotted in the 
figure are the contributions of the individual dust species to the spectrum. 
The solid lines show the contribution of the amorphous silicate
with an olivine stoichiometry (marked in the figure with Ol) and forsterite (Fo). 
The dotted, dashed and dot-dashed lines show the contributions of carbonaceous material, 
ice and iron oxide grains, respectively. The curves of forsterite, carbonaceous, 
ice and iron oxide are offset by -10,-50,-70 and -90 Jy, respectively.
}
\label{fig-halebop}
\end{figure}

Plotted in Fig.~\ref{fig-halebop} is the ISO-SWS spectrum of comet Hale-Bopp, together with our best
fit model. Also shown are the contributions to the spectrum of the individual dust species.
The parameters of the model fit are listed in Table~\ref{hd100546:tbl-3}. 
As has been discussed by previous authors \citep[e.g.][]{hd100546malfait}, the ISO spectrum of Hale-Bopp 
bears a striking resemblance to that of HD\,100546. Our modeling efforts of the spectrum 
of Hale-Bopp will focus on establishing the similarities and differences in dust properties 
between the comet and HD\,100546.

\subsubsection{Grain composition and properties}

In general we find that the dust composition of Hale-Bopp is similar to HD\,100546,
with a few exceptions.
We find no evidence in the spectrum of Hale-Bopp for emission from metallic iron grains as in the case
of HAEBE systems. This could be a spurious result as in the cold (160~K) dust of Hale-Bopp
iron grains have no distinct spectroscopic
feature and their emission could blend with that of other smooth continuum sources, 
for instance from carbonaceous grains. However, it could also imply that metallic
iron grains are really absent, given the suggested high temperature 
formation mechanism of this grain species (see Sect.~\ref{subsec:hd_small_grains}). 
At lower temperatures metallic iron can also react to form iron sulfides or oxides 
\citep[e.g][]{gail1998}. Indeed, we find evidence for emission from these latter components,
in this paper modeled with thermal emission from iron oxide grains, similar to that found
in AB\,Aur and HD\,163296 \citep{abaur}, though less abundant.
The possibility, however, exists that the emission component here attributed to iron oxide
is (partially) due to iron sulfide which has similar emission properties in this wavelength region \citep{keller2000}.

\begin{table}
\begin{center}
\begin{tabular}{@{} c|ccccc @{}}
\hline\noalign{\smallskip}
\hline\noalign{\bigskip}
${\rm M}_{\rm dust}$&\multicolumn{4}{l}{$4.2\cdot 10^{9}$ kg} \\
\multicolumn{5}{l}{}  \\ \hline \\
Dust species&R [AU]&$\rho_0$ [gr cm$^{-3}$]&${\rm M}_{\rm frac}$&a [$\mu$m]\\
\noalign{\smallskip}
Amorph. silicate&2.8--3.0&$3.4\cdot 10^{-29}$&0.64&0.01--10\\
Carbon&2.8--3.0&$1.3\cdot 10^{-29}$&0.23&0.5--10\\
Water Ice&0.6--0.8&$4.0\cdot 10^{-29}$&0.03&15\\
Iron oxide&12.6--12.8&$1.1\cdot 10^{-31}$&0.03&--\\
Forsterite&0.9--1.1&$3.0\cdot 10^{-29}$&0.07&--\\
\noalign{\smallskip}
\hline
\end{tabular}
\end{center}
\vspace{-0.4cm}
\caption[]{Model fit parameters of Hale-Bopp. Listed are the parameters defining the density 
and grain size distribution, the chemical composition and the mass fraction ${\rm M}_{\rm frac}$ 
of the individual dust species. For all dust components we assumed a power law density distribution 
$\rho(r)\propto r^{-1}$ and size distribution $n(a)\propto a^{-2.8}$. The radial extend and grain size
range are given in the table.
\label{hd100546:tbl-3} 
          } 
\end{table}

A component of amorphous silicates and carbonaceous grains similar as in HD\,100546, 
constituting  a major fraction of the dust mass, is seen in Hale-Bopp. 
Interesting to note is the typical grain size of the
cometary dust, which follows a power-law slope $m=2.8$, much larger than in HD\,100546 ($m=2$).
Note that for the forsterite and iron oxide grains we used a CDE shape distribution which 
does not allow for the determination of grain size. 
Spectroscopically, the best fit to the forsterite bands is, like for HD\,100546, achieved by using the
measurements of \citet{servoin1973}. 
Similar uncertainties apply for the derived forsterite mass fraction as
for HD\,100546, discussed in Sect.~\ref{sect:hd_small_Xsil}.
Our best estimates for the effects of a grain size distribution on the derived mass indicates that the
maximum mass fraction of forsterite could be about 
40~\%, i.e. six times higher than in HD\,100546.
A grain size distribution would also result in a higher continuum contribution of the forsterite grains, 
which eliminates the requirement for an additional component in the form of iron oxide.
From the ISO-SWS spectrum it is difficult to determine
the characteristics of the water ice component. We therefore used the results of \citet{lellouch1998},
who found from an analysis of the ISO-LWS spectrum that the average water ice grain 
is 15~\micron.
The mass fraction of the ice is similar to what \citet{lellouch1998} find, however, 
they infer a total dust mass of $1.1 \cdot 10^{11}$~kg, almost two
orders of magnitude larger than we derive. They however, assumed a grain size of
100~\micron\ in deriving the total dust mass. Lowering this to the sizes we find would scale 
their total dust mass to a similar value as found here.

Previous studies \citep[e.g.][]{wooden1999} have also used pyroxenes to model the observations
of Hale-Bopp. However, from the ISO spectrum no evidence can be found for the precence
of such a component. We have therefore not used these dust species in our modeling.

\subsection{Thermal contact of grain species}
\label{sect:hd_thermal_contact}

The lower panel of Fig.~\ref{hd100546:fig-mass2} shows the mass-temperature distribution of Hale-Bopp. 
Contrary to the HD\,100546 system, 
the dust species in Hale-Bopp all have the same mass averaged temperature of approximately 160~K.
The difference in the width of the distribution of the individual dust species results from the assumption
of a grain size distribution for the amorphous dust, giving a broader distribution than 
for the other dust species (forsterite, iron oxide)
where a CDE shape distribution is assumed. 
In order to model the spectrum of comet Hale-Bopp, we found, as
expected, that the bulk of the dust material (i.e. the amorphous
silicates and the carbon, together accounting for 87 percent of
the dust mass) has to be placed at the proper heliocentric 
distance at the time of observation (2.8~AU and distributed over
0.2~AU), giving it the required mass averaged temperature of $\sim$160~K.
To fit the near-IR wavelength range, 
positioning the iron oxide at the correct distance resulted 
in a temperature for this species too high to properly fit the 
spectrum. We determined the best fit temperature for this material 
by positioning it at a larger distance. This yields T $\sim 160$~K.
The same approach was followed to determine the temperature of 
the water ice and the crystalline silicate. Here it was found that
due to their relatively poor light absorption properties
both materials had to be placed closer to the star, but again at 
such a distance that their bulk mass averaged temperature is 160\,K.
A co-spatial distribution of all grain species, complying with the derived mass-temperature 
distribution and the observational constrains on the heliocentric distance, 
can only be obtained if thermal contact between the grains is imposed. 
This result implies that all dust particles are in thermal
contact with one another and that their temperature is essentially
determined by the materials dominating the thermal energy content 
of the dust. The latter are the amorphous silicates and the 
carbonaceous material.
The typical dust temperature we find is much lower than that addopted in previous
studies by \cite{wooden1999}. They find a typical temperature for the
amorphous silicates and crystalline olivine of 245~K, which is $\sim$  85~K higher
than what we find. Also, they suggest a temperature differnce between
crystalline pyroxene grains and other dust species which is contradictory to
our findings.
 
Though the dust species are the same as in 
HD\,100546, Fig.~\ref{hd100546:fig-mass2} shows the fundamental difference 
between the dust in Hale-Bopp and that
in the Herbig Be system: The grains in Hale-Bopp are in thermal contact, those in HD\,100546 are not.
Forcing the grain species in HD\,100546 to be in 
thermal contact, the observed difference in the mass-temperature distribution between the amorphous and
crystalline dust can not be reproduced. This leads to the conclusion that the grains in 
HD\,100546 are homogeneous, separate entities.

\section{Discussion}
\label{sect:hd-discussion}

To summarize, the main results from the analysis of the SED of HD\,100546 are the following:
(1) The SED of HD\,100546 stands out compared to other well studied isolated HAEBE systems 
in the sense that it has a much weaker near-IR excess, but a much stronger mid-IR 
excess. (2) An additional component of
small ($\le$10~$\mu$m) dust grains is present in the circumstellar material surrounding
HD\,100546, not seen in other systems. (3) HD\,100546 has by far the largest crystalline silicate mass 
fraction of all HAEBE stars studied with the ISO satellite. (4) A comparison of the mass-temperature
distribution of the small, spectroscopically identifiable amorphous and crystalline silicates shows
that the bulk of the crystalline silicates has a much lower temperature than that of the amorphous
material. From these results several conclusion concerning the disk geometry and the origin
of the crystalline silicates can be drawn, as we will discuss below.

\subsection{Disk Geometry}
\label{sect:disk_geometry}

As we showed in Sect.~\ref{sect:hd-energy}, the SED of HD\,100546 deviates considerably from that of AB\,Aur.
The SED of the latter can be reproduced with a hydrostatic radiative equilibrium disk
model \citep{dullemond2001}. The unusual luminosity distribution of the circumstellar material
as listed in Table~\ref{tbl-4}, however, can not be reproduced with a model having a similar
disk geometry as used to model AB\,Aur. 
For dust grains to efficiently re-radiate the absorbed stellar light in the mid-IR, 
the typical dust temperature must be in the order of 200~K, which is
the equilibrium temperature of grains in the disk surface layer at a radial distance of about 10~AU.
To reproduce the observed mid-IR luminosity of 0.35~L$_\star$ the relative scale height $H/R$ 
of the disk surface at the radial distance of $\sim$10~AU has to be 0.35 
in order to ``capture'' this amount of starlight. 
This is a factor of two higher than the derived scale height at this radius of the disk surface  in 
the AB\,Aur system \citep{dullemond2001}. 

However, as \citet{dullemond2001} show, if a hole forms in the
disk, the inner edge of the disk will ``puff up'' as a consequence of
the direct irradiation by the central star. The scale height at the
edge will typically increase by a factor of two over the height of a
normal flaring disk without inner hole. Using the same model
parameters as derived for AB\,Aur, but with a hole of 10~AU would
reproduce the required disk scale height and consequently the observed
mid-IR excess of HD\,100546.  However, the region inside 10\,AU cannot
be free of dust since there is still excess emission in the near
infrared.  The material producing this excess must be located at
distances of $\la$1~AU. Instead of an inner hole, the disk should
therefore have a gap starting outside of $\sim 1$\,AU and extending to
about 10\,AU.  The question is then, whether enough stellar radiation
can reach the inner boundary of the outer disk in order to still cause
the increased scale height there.  In the \citet{dullemond2001} model,
the material in the inner regions casts a shadow over parts of the
disk.  However, as we have shown in Sect. 2, the near-IR excess of
HD100546 caused by the inner part of the disk is relatively small, a
disk surface height of $H/R = 0.085$ is geometrically sufficient to
absorb the required 8.5\% of the stellar radiation.  Therefore the
blocking of light by the hot dust is only a small effect in HD100546.
We can now identify the mid-IR component in the ISO spectra with a
``wall'', being the far side of a disk gap directly illuminated by the
star.  The additional dust component of small grains (zone 2, see
Table~\ref{tbl-2}), unique to HD\,100546, should then be the dust
being present in this ``wall'' at $\sim$10~AU.

\subsection{Minearology: the origin of the crystalline silicates}
\label{sect:origin_Xsil}

\begin{figure}[t!]
\centerline{\resizebox{\hsize}{!}{\includegraphics[angle=0]{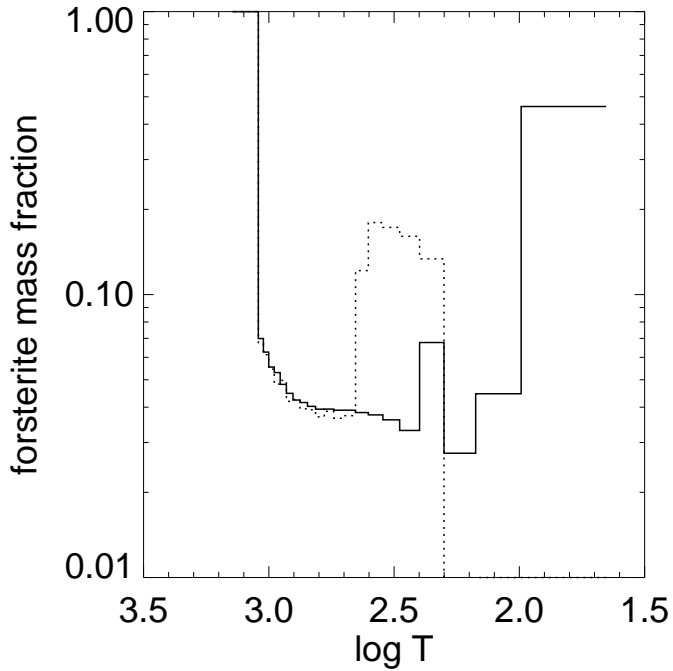}}}
\caption{Relative forsterite mass fraction as a function of temperature. The solid line indicates
our best model fit. The dotted line represents a model where the crystalline and amorphous 
silicates are co-spatial with a constant relative mass fraction as a function of radial distance.
Note that between the mass-temperature and spatial density distributions an opacity effect exists. 
The forsterite grains, having lower opacities in the UV and optical compared to the amorphous
silicate, have a different temperature distribution. This shift in the temperature distribution 
between the amorphous and crystalline dust causes the rise seen around $\sim$300~K for the model
with constant relative mass fraction as a function of radial distance.
}
\label{fig-fraction}
\end{figure}

As was already noted in previous studies, HD\,100546 has the highest known fraction of crystalline 
silicates of all HAEBE stars \citep{hd100546malfait, HerbigOverview, processing}. More puzzling,
the bulk of the crystalline silicates is at low temperatures 
($\sim$ 100 K, see Fig.~\ref{hd100546:fig-mass1}). The question is how to explain both the high
mass fraction as well as the mass-temperature distribution of the crystalline silicates.
Fig.~\ref{fig-fraction} shows the relative forsterite mass fraction,
as plotted in  Fig.~\ref{hd100546:fig-mass1}, relative to the 
total mass in silicates as a function of temperature.
Clearly visible is the rise in the crystalline 
mass fraction at lower temperatures.  A crucial question is which mechanism 
is responsible for this outward increase of the crystalline dust fraction.
This question is closely related to the formation processes of the crystalline silicate. 
Forsterite will form by annealing of amorphous silicate at temperatures above the glass temperature.
At temperatures above 1100~K the annealing time-scale
become much smaller than any dynamical time-scale \citep[e.g.][]{hallenbeck2000}, 
and a fast transformation into crystalline silicates follows. 
This implies that the dust in the very inner parts of the circumstellar disk, 
with temperatures above 1100~K, will be  entirely crystalline up to the evaporation 
temperature of silicate grains at about 1400~K.
Several radial mixing models have been proposed
to transport material outwards to the regime where crystalline silicates are observed
well below the glass temperature. \citet{stevenson1990} predicted that 
the mass fraction of the mixed material would decline outwards following a power law. 
\citet{prinn1990} argued that the neglect of nonlinear
momentum terms in the Stevenson model could underestimate the radial mixing. He predicts that
the mass fraction of the mixed material could almost be constant as a function of radius.
Similar results are found in recent models by \citet{dominique2000} and \citet{gail2001}. 
Plotted in Fig.~\ref{fig-fraction}
(dotted line) is a model with a constant relative mass fraction of forsterite over amorphous silicate
as a function of radial distance, representing the case of the most effective radial mixing predicted.
As can be seen from this figure, outward radial mixing of material from the inner parts 
of the proto-planetary disk can not be the explanation for the observed distribution of 
crystalline silicates.

Also, \citet{processing} show that the
dust composition  producing the 10\micron\ silicate feature, probing grains at the inner parts 
of the proto-planetary disk, is inconsistent with thermal annealing. Clearly  a different mechanism for
the formation of the observed crystalline dust fraction and distribution must be formulated.
Having ruled out radial mixing, this implies that the crystalline silicates must have formed locally, i.e.
further out in the disk where they are presently observed. 
What could have caused this crystallization in the cool outer regions of the disk?
One way of providing enough energy for the silicates to crystallize is to
incorporate them into larger bodies. Radioactive decay of species like $^{26}$Al could cause
the (partial) melting and consequently crystallization of silicates in sufficiently 
large parent bodies \citep[e.g.][]{huss2001}. Afterwards, 
small crystalline dust grains can be produced by collisions between the processed parent
bodies.  If the parent body was large enough to fully
melt during its accretional phase, it will be differentiated.  The
crust will be deprived of siderophile elements like Fe/Co/Ni.  The
composition of the crust will therefore largely be a magnesium-rich
silicate, very similar to what is indicated by observations.
For a collisional cascade of initially large objects to proceed, the orbits must
be sufficiently non-circular to produce large relative velocities. For this one needs many Pluto
sized objects or a large Jupiter type planet \citep[e.g.][]{charnoz2001,kenyon2001}.

An other possible mechanism for {\em in situ} production of
crystalline silicates has been proposed by \citet{harker2002}. They put
forward the possibility that shock waves triggered by gravitational
instabilities could heat small amorphous grains to sufficiently high
temperatures to crystallize.  However, this mechanism still implies
thermal annealing of small dust grains, and the 8.6\,\micron\ feature
of silica should still be produced in this case.

\subsection{Proposed scenario}
\label{sect:scenario}

As we have discussed in the previous two subsections, there are two things that
set HD\,100546 apart from other HAEBE stars, being its disk geometry and its
mineralogy. We propose a scenario which can explain and connect both deviations.
In this section we will speculate that a massive planet has formed in
the disk at a distance of about 10~AU from the star, and that this
planet is responsible for the striking difference in both the
mass-temperature distribution and in the chemical composition and
crystallographic structure of the dust around HD\,100546.

A massive planet will be able to strongly influence planetesimals 
and the structure of the gas disk
within its Hill sphere which is given by \citep{hill1878} 
\begin{equation}
\label{eq:1}
h=\left( \frac{M_{\rm pl}}{3M_{\star}} \right)^{1/3} A
\end{equation}
where $A$ is the semi-major axis of the planets orbit, $M_{\rm pl}$ is
the mass of the planet and $M_{\star}$ is the mass of the star.  
\citet{lin1993} showed that a planet will open a gap in the disk
if its Hill sphere equals or exceeds the vertical disk scale height.
Assuming that before the formation of a planet the disk was very similar 
to that of AB\,Aur, the relative disk scale height at 10~AU
would have been $H/R=0.045$ \citep{dullemond2001}. Substituting this into Eq.~(\ref{eq:1}), 
taking into account that the far side of the gap is ``puffed up'' by a factor of about two
(see Sect.~\ref{sect:hd-energy} and Sect.~\ref{sect:disk_geometry}),
would yield a planetary mass of $M_{\rm pl} \ge 5.6 M_{\rm J}$, indicating a Jupiter type planet.
A formation distance of $\sim$10~AU for a gas giant seems reasonable to assume,
in particular since the temperatures in the disk around a late type B star will
be similar at 10~AU to the conditions at 5~AU in the early solar
system. It has been suggested that the \emph{snow-line}, i.e. the line
where water ice can condense and remain stable in the disk, has
enhanced the densities in the solar disk locally to speed up the
accretion of Jupiter \citep{1988Icar...75..146S}. This ice condensation
zone for the HD\,100546 system is located at about 10~AU.
Also, the time scale for the 
formation of a giant planet in the disk region is 
a few Myrs \citep{pollack1996,hersant2001}, which can easily be accounted for 
given the estimated age of HD\,100546 of $>$10~Myr \citep{mario1998}.
A planet opening a gap in the disk would also be a natural explanation
for the lower near-IR excess of HD\,100546 compared to AB\,Aur. Such a gap
could cause the depletion of the inner disk by starvation of material
from the outer disk, though \citet{1999ApJ...526.1001L} argue that for
planets of $\sim$M$_{\rm J}$ still a net flow through the gap can exist
not completely clearing the inner region.

\begin{figure}
\centerline{\resizebox{\hsize}{!}{\includegraphics[angle=-90]{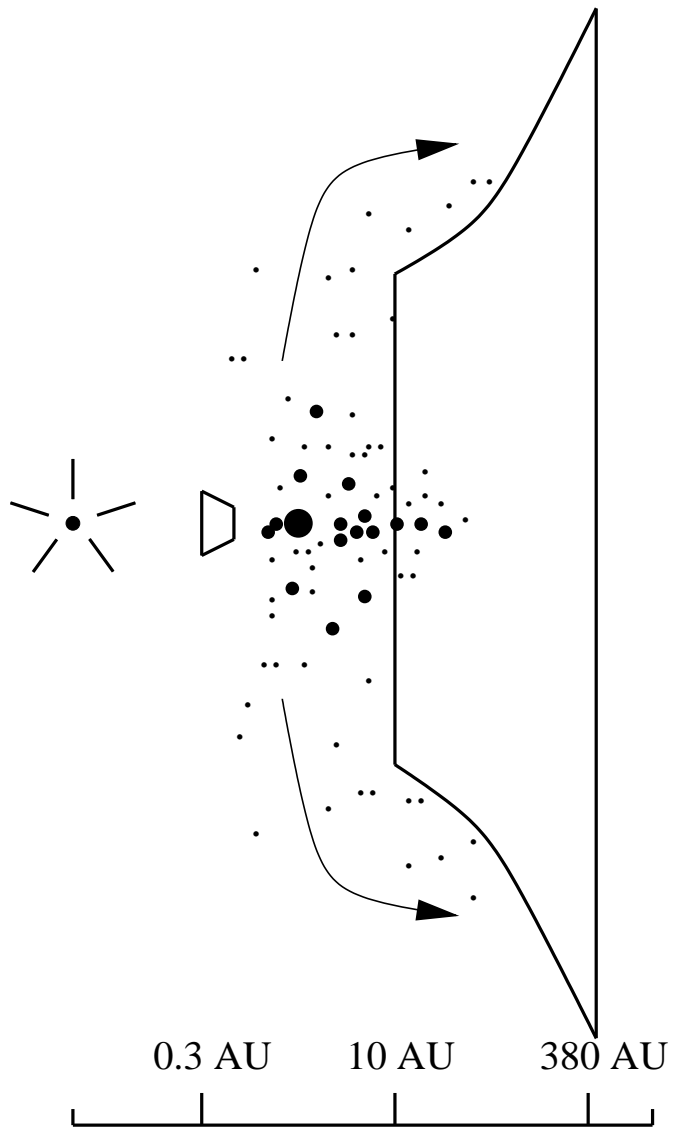}}}
\caption{Schematic representation of the HD\,100546 system. A giant planet has
formed a gap in the flaring disk at 10~AU, causing an increase of the disk scale height.
A collisional cascade induced by the giant planet is producing dust and throwing it up out
of the disk mid-plane. The small  crystalline grains above the disk are 
blown away by radiation pressure and rain down  further out at a few hundred AU on the surface of the disk
(arrows).
} 
\label{fig-disk}
\end{figure}

As briefly mentioned in Sect.~\ref{sect:origin_Xsil}, the formation
mechanism for the crystalline silicates could be a collisional cascade
of planetesimals.  A Jupiter mass planet can easily ``heat up''
the orbital elements of a swarm of planetesimals through gravitational
stirring, causing sufficiently high relative velocities for collisions
to be destructive.  Indeed, \citet{stern2001} show that local
collisional erosion of planetesimals may be a more important process
for planetesimals near a giant planet than ejection to the Oort
cloud. An important question is now, how much reprocessed dust can
be expected from a collisional cascade of such bodies.
\cite{dohnanyi} has shown that the collisional cascade leads to a size
distribution $f(a)\propto a^{-3.5}$ or $f(m)\propto m^{-11/6}$ in
which the mass is dominated by the largest particles.  If the
observations measure all particles up to a certain size limit
$a_1\approx10$~\micron\ and the maximum size of planetesimals involved
in the collisional cascade is $100$~km, then one finds $m_{\rm
  vis}/m_{\rm tot}=10^{-5}$.  For the measured amount of crystalline
dust of $m_{\rm vis}=1.3\times10^{-7}$~M$_{\odot}$, one finds a total
mass of $m_{\rm tot}=1.3\times10^{-2}$~M$_{\odot}$.  This is more than
there is available in the disk.  Therefore, the observed dust cannot
be the steady state solution of a collisional cascade.


However, the steady state solution assumes that small particles are
constantly removed at the low end of the size distribution.  If the
disk around the star were already gas-free, particles smaller than
about 10~$\mu$m in size would be blown away by radiation pressure
\citep[e.g.][]{Arty-88}.  However, the disk around HD 100546 very likely is not gas
free.  Even if large bodies can move around
freely on purely Keplerian orbits, the motions of small grains will be
slowed down.  This will keep small grains in the disk, and it will
also drastically reduce the collision velocities between small grains.
Therefore, small grains are no longer destroyed in collisions.
Grains which are small enough will accumulate in the disk.
The density of these grains will not be in
collisional equilibrium.  It will instead mirror the entire mass of
larger bodies destroyed earlier. This greatly softens the mass
constraint we derived above.  If the small grains accumulate, only of
order 1 $M_{\rm Earth}$ of crystalline material has to be set free in
collisions, a number we can comfortably account for.

We note that the crystalline dust in zone~2 of the small grain component is distributed in
the model fit over a larger range than the amorphous dust.  One way in
which this could be realized is that small dust grains produced by the
collisional cascade will be driven out by radiation pressure.  Because
the disk is flared, the grains will not be able to leave the system,
but will ``land'' further out on the disk surface.  There they will
enrich the super-heated disk surface layer which is known to be
responsible for the features in the spectra of
irradiated flaring disks \citep{CG}.  This
redistribution of processed dust is similar to the model proposed by
\citet{1996Sci...271.1545S} for the origin of Chondrules and 
calcium-aluminum rich inclusions (CAIs) in
meteorites.  In the \citeauthor{1996Sci...271.1545S} model, grains are
lifted off the disk by the jet which is leaving the disk very close to the
star in the active accretion phase of the disk.  The grains then are
heated up to melting temperatures by the stellar light and transported
to larger distances by radiation pressure.  In what we propose, the
grains are produced by fragmentation of larger bodies, and are
scattered out of the disk.  Small grains for which radiation pressure
is dynamically important will be driven away from the star and by a
combination of gravity and disk geometry land further out on the disk
surface.  This is the location where we see these grains today.
A schematic representation of the above sketched disk structure and processes
is plotted in Fig.~\ref{fig-disk}.

An alternative scenario for the {\em in situ} formation of the crystalline silicates is
annealing by nebular shocks. \citet{harker2002} argue that local gravitational 
instabilities, likely to exist in protoplanetary disk beyond 5 to 10~AU, could 
cause shocks. These shocks could sufficiently heat small ($\le 1\mu$m)
amorphous grains in order  
to crystallize them. Interestingly, a developing planet, opening a gap in the disk, 
would produce shock waves through tidal interaction with the disk \citep[e.g.][]{lin1980,lin1993}.
However, the thus generated shocks should radially extend over the very wide observed region of the
forsterite grains. It remains to be seen whether the gas densities are high enough in regions
extending far beyond 10~AU for this mechanism to be effective.

The consequence of the above proposed scenario is that Hale-Bopp, given the very similar dust composition, 
may contain material formed by collisions of bodies formed earlier on in the history of the Solar system.
The thermal contact of the dust species in Hale-Bopp
implies that we are not witnessing the destruction of Hale-Bopp type objects in the HD\,100546 system,
but the creation of the material similar to that which
formed Hale-Bopp before the dust grains coagulated.
Interestingly this would place the formation of Hale-Bopp after the formation of one
or more of the gas planets in our solar system.

\section{Summary}
\label{sect:hd-summary}

We can summarize the results of our analysis of the ISO-spectra of HD\,100546 and comet Hale-Bopp
as follows:

\begin{enumerate}

\item
The ISO/SWS spectra of HD\,100546 and comet Hale-Bopp can be
fitted with very similar dust compositions and properties. In
both cases, the bulk of the material is dominated by amorphous
silicates with carbonaceous grains as the second most important
constituent. In addition, significant amounts of forsterite, 
crystalline water ice, metallic iron and iron sulfide are present. 

\item
There is one major difference in the dust properties of the
two investigated objects. The chemical constituents
of the dust in Hale-Bopp are found to be in thermal contact
while this is not the case for HD\,100546.

\item
The fraction of crystalline silicates, i.e.
forsterite and/or very iron poor olivine, in 
HD\,100546 increases with decreasing temperature, i.e. with
increasing radial distance from the central star. Such a
distribution is not in agreement with predictions by radial
mixing models. In such models the crystals are formed by
thermal annealing in the very inner parts of the disk and are
subsequently transported out and mixed with amorphous material.

\item
HD\,100546 shows a component of small grains ($<$ 10 $\mu$m)
with temperatures of about 200 K that emits in the mid-IR
and that is not present in the spectra of other
Herbig Ae/Be stars.

\item
At about the distance where the bulk dust temperature is
$\sim$ 200 K, i.e. at $\sim$ 10 AU, the vertical extent of
the disk must be in excess of that expected in a standard flaring
disk model. This is required to explain the large fraction
of stellar light intercepted and re-emitted by grains emitting
in the mid-IR.

\item
On the basis of the above three results, we propose the following
scenario: 
the crystalline dust
is produced by collisional destruction of differentiated objects.
These collisions are probably induced by gravitational interaction
with a proto-Jupiter. The grain component emitting at $\sim$ 200 K
may be identified by the dust expected to be produced in the
collisional cascade that is thought to occur when large bodies collide.
This predominantly crystalline material may be driven out by
radiation pressure and ``land'' further out on the (flaring) disk 
surface. The gap cleared out by the proto-Jupiter allows direct
stellar light to produce an extended rim or wall at the far side of the
gap. It is thought to be this rim that intercepts the radiation 
responsible for the strong mid-IR component in the spectrum of 
HD\,100546.

\item
The similarity in spectral appearance between HD\,100546 and
Hale-Bopp implies that similar processes in the early history of 
our solar system may have been relevant for the formation of 
Hale-Bopp. The comet may have assembled from debris dust produced
through collisions of differentiated objects.

\end{enumerate}

\bibliographystyle{aa}
\bibliography{H3603}

\end{document}